\documentclass[aps,prd,multicol]{revtex4}

\begin{document}
\title{Neutrino mixing in a left-right model}
\author{J. A. Martins Sim\~oes, J. A. Ponciano\\
Instituto de F\'\i sica,\\
Universidade Federal do Rio de Janeiro, RJ, Brazil \\}
\begin{abstract}
We study the mixing among different generations of massive neutrino fields in a $SU(2)_L \times SU(2)_R \times U(1)_Y$ gauge theory which includes Majorana and Dirac mass terms in the Yukawa sector. Parity can be spontaneously broken at a scale $v_R\simeq 10^3-10^4$ GeV. We discuss about possible candidates for the Yukawa coupling matrices and we found that the model can accommodate a consistent pattern for neutral fermion masses as well as neutrino oscillations. The left and right sectors can be connected by a new neutral current.
\end{abstract}
\maketitle
\section{Introduction}
The increasing experimental evidence on neutrino oscillations and non zero masses \cite{1} brings new light in some deep physical questions. The present value for the neutrino masses 
is consistent with a see-saw mass generation description involving a large mass scale.
This suggests that grand unified theories have an important role in the neutrino mass spectrum. If this is the case, we still have many other points to clarify since the standard model has a relatively large number of input parameters and properties. In a recent work \cite{2,3}, an extended model was proposed in other to clarify two of these points; the origin of parity breaking and the small value of the charged lepton mass spectrum relative to GUTs scales.
\par
 One possible way to understand the left-right asymmetry in weak interactions is to enlarge the standard model into  a left-right symmetric structure and then, by some spontaneously broken mechanism, to recover  the low energy asymmetric world. Many models were developed, based on grand unified groups \cite{4}, superstring inspired models \cite{5}, a connection between parity and the strong CP problem \cite{6}, left-right extended standard models \cite{7}. All these approaches imply  the existence of some new intermediate physical mass scale, well bellow the unification or the Planck mass scale. Left-right models starting from the gauge group \break
$SU(2)_{L}\otimes SU(2)_{R}\otimes U(1)_{B-L}$ were developed by many authors \cite{8} and are well known to be consistent with the standard  $SU(2)_{L}\otimes U(1)_Y$. However, for the fermion mass spectrum there is no unique choice of the Higgs sector that can reproduce the observed values for both charged and neutral fermions, neither the fundamental fermionic representation is uniquely defined. 
\par
In ref.2, a left-right model with mirror fermions and a particular choice for the Higgs sector was proposed, leading to a new see-saw mass relation for both neutral and charged leptons. In the present paper we extend the model to three families and study the consequences for neutrino masses and oscillations. In section II we briefly review the model for completeness. In sections III and IV we present the charged and neutral lepton mass spectrum respectively, obtained from possible candidates for the Yukawa coupling matrices. In section V we discuss, from the neutrino mixing matrix, the consequences for neutrino oscillations; in section VI we present the main phenomenological consequences and, finally, our conclusions are given in the last section.

\section{The model}
In this section we will briefly revise the most important features of the model. Details can be found on ref.2 and 3.

We will analyze the neutral fermion masses within the framework of a theory based on the gauge structure $SU(2)_L \times SU(2)_R \times U(1)_Y$ with coupling constants $g_L$,$g_R$ and $g$. 

An explicit realization of the model is provided by specifying its Higgs particle and fermion content. The first family assignment of standard and exotic fermions to the $SU(2)_L \times SU(2)_R \times U(1)_Y$ representations is as follows:
\begin{eqnarray}
\begin{array}{cccc}
l_L=\left(\begin{array}{c}
\nu\\ 
e
\end{array}\right)_L&
(1/2,0,-1),&
L_R=\left(\begin{array}{c}
N\\
E
\end{array}\right)_R&(0,1/2,-1),\\ 
\nu_R &(0,0,0), & N_L &(0,0,0), \\
e_R &(0,0,-2), & E_L &(0,0,-2).
\end{array}
\end{eqnarray}
The electric charge operator is defined in terms of the generators $T_L$, $T_R$ and Y of $SU(2)_L$, $SU(2)_R$ and $U(1)$ respectively.
\begin{equation}
Q= (T_{3L}+T_{3R}+Y)
\end{equation}
A general choice of the Higgs sector includes a Higgs field $\Phi$ in the mixed representation $(1/2,1/2,0)$ and two Higgs doublets 
\begin{equation}
\chi_L=\left(\begin{array}{c}
\chi_L^+\\
\chi_L^0
\end{array}\right),\quad
\chi_R=\left(\begin{array}{c}
\chi_R^+\\
\chi_R^0
\end{array}\right),
\end{equation}
with transformation properties
\begin{equation}
(1/2,0,1)_{\chi_L}, \qquad (0,1/2,1)_{\chi_R}. 
\end{equation}
The breakdown of $SU(2)_L\times SU(2)_R \times U(1)_Y$ down to $U_{em}(1)$ is realized through a non trivial pattern of vacuum expectations values for the Higgs fields, namely,
\begin{equation}
<\chi_{L}>=\left(\begin{array}{c}
0\\
v_L\end{array}\right), \quad
<\chi_{R}>=\left(\begin{array}{c}
0\\
v_R\end{array}\right), \quad 
<\Phi>=\left(\begin{array}{cc}
k&0\\
0&k'\end{array}\right).
\end{equation}
Higgs doublets are responsible for the gauge boson and fermion masses. To the Higgs sector we add two new Higgs singlets, one coupled to Dirac mass terms -$S_D$- and the other coupled to Majorana mass terms - $S_M$.

At present there are several indications in favor of nonzero neutrino mass and mixing between families coming from solar and atmospheric neutrino data \cite{Fornengo}. Neutrinos are predicted to be Majorana particles in many extensions of the standard model containing neutrinos with nonzero masses. Here we will do so and allow Majorana mass terms within the Yukawa sector of the lagrangian. For the first family we have
\begin{eqnarray}
\mathcal{L}_M&=& f\left[\overline{l}_L \chi_L e_R + \overline{L}_R \chi_R E_L+\overline{l}_L\tilde{\chi}_L\nu_R+\overline{L}_R\tilde{\chi}_R N_L\right] + \\ \nonumber
&+& f'\left[\overline{l}_L\tilde{\chi}_LN^c_L+\overline{L}_R\tilde{\chi}_R\nu^c_R\right]
+ f''\left[\overline{l}_L\phi L_R\right] +\\ \nonumber
&+& g S_M\left[\overline{N^c}_LN_L+\overline{\nu^c}_R\nu_R\right]+ g'S_D\overline{\nu}_RN_L
+g''S_D\overline{e}_RE_L. 
\end{eqnarray}
The generalization to three families is straightforward.
Notice that the inclusion of Majorana terms spoils the invariance with respect to any global gauge transformation so that there is no conserved leptonic charge (see for example Ref. \cite{Bilenky}).

Fermions masses arise after spontaneous symmetry breaking of the gauge 
structure $SU(2)_L\times SU(2)_R \times U(1)_Y$ down to $SU(2)_L\times 
U(1)$. For the charged and neutral sectors the mass lagrangians are, 
respectively
\begin{eqnarray}
\mathcal{L}_{M,c}=f\left[v_L\overline{e}_Le_R +v_R 
\overline{E}_R E_L\right] 
+g''s_D\overline{e}_RE_L+f''k'\overline{e}_LE_R+ H.C.,
\end{eqnarray}
and, 
\begin{eqnarray}\label{nyukawa}
\mathcal{L}_{M,n}&=&f\left[v_L\overline{\nu}_L\nu_R+v_R\overline{N}_RN_L\right]+
f'\left[v_R\overline{N}_R\nu^c_R+
v_L\overline{\nu}_LN^c_L\right]+f''k\overline{\nu}_LN_R\\ \nonumber
&+&gs_M\left[\overline{N^c}_LN_L+\overline{\nu^c}_R\nu_R\right]+ g's_D\overline{\nu}_RN_L+ H.C. .
\end{eqnarray}

One of the main points of the mirror left-right model is the presence of the term $g''s_D\overline{e}_RE_L$ in the charged lepton mass matrix. This term will imply a see-saw mass relation for the charged sector. We have then a natural mechanism to explain small charged lepton masses in a large unified mass scale.
\par

In this model where CP violation is not taking into consideration and therefore the couplings $f$, $f'$, $g$, $g'$, $g''$ and $h'$ are $3 \times 3$ real matrices.

In matrix form, taking $k=k'=0$, the charged sector reads
\begin{eqnarray}\label{matrixcharged}
\mathcal{L}_{M,c}&=&\overline{\psi}M_{c}\psi, \\ \nonumber
&=&\left(\overline{f}_{L},\overline{F}_R,\overline{F}_L,
\overline{f}_R\right)
\left(\begin{array}{cccc}
0 & 0 & 0 &fv_L\\
0 & 0 & fv_R &0 \\
0 & fv_R & 0 & g''s_D \\
fv_L & 0 & g''s_D & 0 
\end{array} \right)
\left(\begin{array}{c}
f_L\\
F_R\\
F_L\\
f_R
\end{array} \right).
\end{eqnarray}

On the other hand, for the neutral sector it is convienient to introduce the self-conjugated fields defined, for each family, as 
\begin{eqnarray}\label{majoranafields}
\chi_{\nu}&=&\nu_L+\nu_L^c \\ \nonumber
w_N&=&N_R+N_R^c\\ \nonumber
\chi_{N}&=&N_L+N_L^c\\ \nonumber
w_{\nu}&=&\nu_{R}+\nu_{R}^c
\end{eqnarray}
In terms of the new fields, equations (\ref{nyukawa}) may be rewritten as
\begin{eqnarray}\label{matrixneutral}
\mathcal{L}_{M,n}&=&\overline{\xi}M_n\xi \\ \nonumber
&=&\left(\overline{\chi}_{\nu},\overline{w}_N,\overline{\chi}_N,
\overline{w}_{\nu}\right)
\left(\begin{array}{cccc}
0 & 0 & f'v_L &fv_L/2\\
0 & 0 & fv_R/2&f'v_R \\
f'v_L & fv_R/2 & gs_M & g's_D/2 \\
fv_L/2 & f'v_R & g's_D/2 & g''s_M 
\end{array} \right)
\left(\begin{array}{c}
\chi_{\nu}\\
w_N\\
\chi_N\\
w_{\nu}
\end{array} \right).
\end{eqnarray}
The mass matrices show the following block structure with different mass 
scales
\begin{equation}\label{seesaw}
M=\left(\begin{array}{cc}
0 & M_{LR} \\
M^t_{LR} & M_S \end{array}\right)
\end{equation}
where $M_{LR}$ and $M_S$ are $n\times n$ matrices verifying $det(M_{LR})\ll 
det(M_S)$.

In view of the see saw structure ($det(M_{LR})\ll det(M_S)$), mass matrices can 
be driven to a block diagonal form by expanding in power series of $M_{LR}M_S^{-1}$ \cite{Valle}.
This results in a $2n \times 2n$ light fermion mass matrix and $2n \times 2n$ heavy one given by
\begin{equation}
M^{(light)}\simeq - M_{LR}^t M_S^{-1} M_{LR}, \qquad M^{(heavy)}=M_S
\end{equation}
respectively.

\section{Charged fermion masses}
In order to obtain the fermion masses we need explicit textures for the coupling matrices in Eqs.  (\ref{matrixcharged}) and (\ref{matrixneutral}).
Mixing between families in the charged sector is phenomenologically disfavored and thus the coupling matrices $f$ and $g''$ can be chosen diagonal.
Taking $f=diag(1,1,1)$ and $g''=diag(\lambda_1,\lambda_2,\lambda_3)$ we obtain for each family a light charged fermion, with mass eigenvalue $m_i=v_Lv_R/\lambda_is_D$, and a heavy one with  eigenvalue $M_i=\lambda_is_D$.

Flavor left handed and right handed fields $\psi$ are connected to the physical fields $\eta$ by means of an orthogonal transformation, that is
\begin{equation}
\psi_{j}=\sum_{k=1}^{4n}V_{jk}\eta_{k}
\end{equation}  
where $\eta$ is the column matrix formed by the mass fields and $n$ the number of families into consideration. 

Explicitly, the mixing matrix $V$ in the one family case is
\begin{equation}
V=\left(\begin{array}{rrrr}
1&1&0&0\\
1&-1&0&0\\
0&0&1&1\\
0&0&1&-1
\end{array}\right)
\end{equation}
From this matrix we can recover the Dirac structure of charged leptons by suitable rotations. 
The generalization to the three family case is easily done.

The three parameters $\lambda_i$ in $g''$ allow us to recover the standard charged fermion spectrum in a simple way. 
\par
For the light fermions we have
\begin{equation}
m_i=\frac{v_Lv_R}{\lambda_is_D} 
\end{equation}
Fixing the vacuum parameter $v_L$ equal to the Fermi scale $v_{Fermi}$, we obtain the following constraints
\begin{equation}
v_R/{\lambda_1s_D}\simeq  10^{-6},\qquad v_R/{\lambda_2s_D}\simeq 10^{-3}, \qquad v_R/{\lambda_3s_D}\simeq 10^{-2}.
\end{equation}
Consequently, for $v_R\simeq 10^3 -10^4\; GeV$ we have the following spectrum for the heavy sector
\begin{equation}
M_1=\lambda_1 s_D=10^9-10^{10}\; GeV, \quad M_2=\lambda_2 s_D=10^6-10^7\; GeV, \quad M_3=\lambda_3 s_D=10^5-10^6 GeV.
\end{equation}
Taking the vacuum expectation value of the Higgs scalar $s_D$ at the mass scale $10^{10}$ GeV, then the coupling parameters $\lambda_i$ are fixed to
\begin{equation}
\lambda_1=1, \qquad \lambda_2=10^{-3}, \qquad \lambda_3=10^{-4}. 
\end{equation} 
\section{Neutral fermion masses}
The mass lagrangian corresponding to the neutral sector contains Dirac and Majorana mass 
terms built up from the inclusion of right handed neutrino fields and their mirror partners. In this framework, the description on the phenomenological neutrino mass matrix will differ from the most familiar schemes on three neutrino mixing found in the literature \cite{Barger}. 
In equation (\ref{matrixneutral}), the Dirac mass terms arise from the off-diagonal submatrices of the blocks $M_{LR}$ and $M_S$, while the Majorana terms arise from the diagonal ones. As we mention before, the difference between $M_{LR}$ and $M_S$ mass scales ensures the see-saw mechanism for the neutral sector.
We still have to choose suitable candidates for the textures of the coupling matrices. There are many possibilities that can be compatible with the present experimental status on neutrino masses and oscillations.

For the Dirac mass terms we chose diagonal couplings. The simplest choice is to take $f$ and $g'$ equal to the unity.  

An important point in the left-right symmetric model comes from the Majorana mass terms. Since Majorana fields are completely neutral and therefore, are all physically equivalent, it is a natural requirement that all Yukawa couplings are to be taken equal. This corresponds to taking democratic textures for the couplings $f'$, $g'$ and $g''$, that is
\begin{equation}
f'=g'=g''=\rho\left(\begin{array}{ccc}
1&1&1\\
1&1&1\\
1&1&1
\end{array}\right),
\end{equation}
where $\rho$ may be set equal to $1$ for simplicity.
The analytic expression for $M^{(light)}$ is now
\begin{equation}\label{massdemocratic}
M^{(light)}=\frac{v_R^2}{s_M}\left(\begin{array}{cccccc}
\frac{1}{4}+3w^2&2w^2&3w^2&\frac{1}{2}+\frac{1}{2}w^2&\frac{1}{2}+\frac{1}{2}w^2&\frac{1}{2}+\frac{1}{2}w^2\\
3w^2&\frac{1}{4}+3w^2&3w^2&\frac{1}{2}+\frac{1}{2}w^2&\frac{1}{2}+\frac{1}{2}w^2&\frac{1}{2}+\frac{1}{2}w^2\\
3w^2&3w^2&\frac{1}{4}+3w^2&\frac{1}{2}+\frac{1}{2}w^2&\frac{1}{2}+\frac{1}{2}w^2&\frac{1}{2}+\frac{1}{2}w^2\\
\frac{1}{2}+\frac{1}{2}w^2&\frac{1}{2}+\frac{1}{2}w^2&\frac{1}{2}+\frac{1}{2}w^2&3+\frac{1}{4}w^2&3&3\\
\frac{1}{2}+\frac{1}{2}w^2&\frac{1}{2}+\frac{1}{2}w^2&\frac{1}{2}+\frac{1}{2}w^2&3&3+\frac{1}{4}w^2&3\\
\frac{1}{2}+\frac{1}{2}w^2&\frac{1}{2}+\frac{1}{2}w^2&\frac{1}{2}+\frac{1}{2}w^2&3&3&3+\frac{1}{4}w^2
\end{array}\right)
\end{equation}
where $w$ is defined as $w\equiv v_L/v_R$.

The interaction fields $\xi\equiv (\chi_{\nu},w_{N})^t$ are related to the physical ones $\eta \equiv (\nu, N)^t$ by means of the orthogonal transformation $\xi=U\eta$,
\begin{equation}\label{orthogU}
\left(\begin{array}{c}
\chi_{\nu i}\\
w_{N i}\end{array}\right)=U
\left(\begin{array}{c}
\nu_i\\
N_i
\end{array}\right)
\qquad i=1,2,3
\end{equation}
The $2n\times 2n$ orthogonal matrix $U$ is determined by requiring
\begin{equation}
U^tM^{(light)}U=diag(m_1,m_2,\ldots ,m_6)
\end{equation}
where $m_k$ are the eigenvalues of $M^{(light)}$ and correspond to the spectrum of the light sector.
Explicitly, we found
\begin{equation}\label{demoU}
U=\left(\begin{array}{cccccc}
\sqrt{\frac{12}{37}}+O(w^2)&0&0&\sqrt{\frac{1}{111}}+O(w^2)&-\sqrt{\frac{2}{3}}& 0\\
\sqrt{\frac{12}{37}}+O(w^2)&0&0&\sqrt{\frac{1}{111}}+O(w^2)&
\frac{1}{\sqrt{6}}&\frac{1}{\sqrt{2}}\\
\sqrt{\frac{12}{37}}+O(w^2)&0&0&\sqrt{\frac{1}{111}}+O(w^2)&\frac{1}{\sqrt{6}}& -\frac{1}{\sqrt{2}}\\
-\sqrt{\frac{1}{111}}+O(w^2)&-\sqrt{\frac{2}{3}}&0&\sqrt{\frac{12}{37}}+O(w^2)&0& 0\\
-\sqrt{\frac{1}{111}}+O(w^2)&\frac{1}{\sqrt{6}}&\frac{1}{\sqrt{2}}&\sqrt{\frac{12}{37}}+O(w^2)&0 &0\\
-\sqrt{\frac{1}{111}}+O(w^2)&\frac{1}{\sqrt{6}}&-\frac{1}{\sqrt{2}}&\sqrt{\frac{12}{37}}+0(w^2)&0& 0\\
\end{array}\right) 
\end{equation}

The neutrino fields are labeled $\nu$ or $N$ according to their characteristic mass scales $v_L^2/s_M$ or $v_R^2/s_M$, respectively. 
The spectrum of light Majorana neutrino masses is

\begin{eqnarray}\label{spectrum}
m_{\nu_1}= \frac{1}{4}\frac{v_L^2}{s_M}, \qquad m_{\nu_2}=\frac{1}{4}\frac{v_L^2}{s_M},\qquad m_{\nu_3}\simeq \frac{1225}{148} \frac{v_L^2}{s_M} \\ \nonumber
m_{N_1}= \frac{1}{4}\frac{v_R^2}{s_M},\qquad m_{N_2}=\frac{1}{4}\frac{v_R^2}{s_M},\qquad m_{N_3}\simeq \frac{37}{4}\frac{v_R^2}{s_M}.
\end{eqnarray}

It is interesting to notice that the model leads naturally to a hierarchical mass spectrum, with  different square mass scales. As we will see in the next section, this feature is essential if the model is to account for the mass pattern coming from neutrino oscillation data. 

From Eq. \ref{spectrum} we can also redefine the six Majorana fields in terms of two Dirac and two Majorana neutrino fields

The main theoretical constraints on neutrino masses come from cosmological considerations related to typical bounds on the universe mass density and its lifetime.
Specifically, the cosmological bound follows from avoiding the overabundance of relic neutrinos. 
For neutrinos below $\simeq 1\; MeV$ the limit on  masses for Majorana type neutrinos is \cite{Turner}
\begin{equation}\label{cosmology}
\sum_{\nu} m_{\nu}\leq 100\Omega_{\nu}h^2\; eV \simeq 30\; eV
\end{equation}
where $\Omega_{\nu}$ is the neutrino contribution to the cosmological density parameter, $\Omega$, defined as the ratio of the total matter density to the critical energy density of the universe and the factor $h^2$ measures the uncertainty in the determination of the present value Hubble parameter $h$. The factor $\Omega h^2$ is known to be smaller than 1.

In Eq.(\ref{cosmology}) the matter component represented by the factor $\Omega_{\nu}h^2$ was chosen smaller than $0.3$, according to reference \cite{Olive} , in order to obtain an age of the Universe $t \geq 12$ Gyears.
 
From (\ref{spectrum}), the sum of neutrino masses satisfy 
\begin{equation}
\sum_i^6 m_i \lesssim 10 \frac{v_R^2}{s_M},
\end{equation}
so that the cosmological criterium (\ref{cosmology}) is verified if
\begin{eqnarray}
\frac{v_R^2}{s_M}\lesssim 10 \Omega_{\nu} h^2\;eV.
\end{eqnarray}
This constrains the breaking scale $s_M$ to be  $s_M\gtrsim 10^{15}\;GeV$ when $v_R$ is fixed at $v_R\simeq 10^3\;GeV$.
 
\section{Oscillations of neutrinos}
The oscillations in neutrino beams are one of the most fundamental consequences of neutrino mixing. Experimental results concerning a two-generation transition are quoted in terms of $\Delta m^2=m^2_2-m^2_1$, and the mixing angle.   
We will see in this section that the model presented previously yields satisfactory results for the democratic texture of the Majorana terms coupling matrices when we fix $v^2_R/s_M\simeq 10^{-2}eV$. 

Taking into account the orthogonality of the mixing matrix, the probability of transition $\nu_{\alpha}\rightarrow \nu_{\beta}$ between two generations $\alpha$ and $\beta$ is 

\begin{equation}\label{probability}
P(\nu_{\alpha}\rightarrow \nu_{\beta})=\arrowvert \delta_{\alpha\beta}+\sum_i^{2n} U_{\alpha i}U_{\beta i} \exp (-i \Delta m^2_{i1} L/2E -1)\arrowvert ^2
\end{equation}
where $L\simeq t$ is the distance between neutrino source and neutrino detector and $E$ is the neutrino energy.

Notice that as a general feature of the transition probability, neutrino oscillations can be observed whenever the condition $\Delta m_{i1}L/E \sim 1$ is satisfied.

Specifically, considering the model in question supplemented by the democratic texture input,
we obtain for the transition $\nu_e\rightarrow \nu_{\mu}$
\begin{eqnarray}
P(\nu_e\rightarrow \nu_{\mu})&=&|U_{e4}U_{\mu 4}\exp(-i\Delta m^2_{41}-1)+U_{e5}U_{\mu 5}\exp(-i\Delta m^2_{51}-1)|^2,
\end{eqnarray}
where the explicit values of the matrix elements $U_{ei}$ are given in Eq. (\ref{demoU}).

In first approximation we neglect $|U_{e4}U_{\mu 4}|=1/111$ in front of $|U_{e5}U_{\mu 5}|=1/3$, yielding to the simpler expression
\begin{eqnarray}\label{emu}
P(\nu_e\rightarrow \nu_{\mu})&=&|U_{e5}U_{\mu 5}\exp(-i\Delta m^2_{51}-1)|^2 \\ \nonumber
&=&\frac{1}{2}4|U_{e5}|^2|U_{\mu 5}|^2\left(1-\cos \Delta m^2_{51} \frac{L}{2E}\right)
\end{eqnarray}
and therefore the amplitude of the probability mixing and the relevant scale of mass are
\begin{equation}\label{solarsol}
4|U_{e5}|^2|U_{\mu 5}|^2=\frac{4}{9}, \qquad\Delta m^2=\Delta m^2_{51}\simeq \frac{1}{16} \frac{v_R^4}{s_M^2}. 
\end{equation}

Recent solar neutrino oscillations results (SNO) strongly favor the large mixing angle Mikheyev-Smirnov-Wolfenstein (MSW) solar solution at the scale 
\begin{equation}
\Delta m^2_{sol} \simeq 10^{-5} eV^2
\end{equation}

Replacing this value in Eq.(\ref{solarsol}) and choosing $v_R\simeq 10^3\;GeV$, we found that in the left-mirror model the singlet breaking scale should be fixed $s_M\simeq 10^{17}\;GeV$ in order to recover the solar neutrino experimental results. 

We now turn our attention to the $\nu_{\mu}\rightarrow \nu_{\tau}$ transition. In this case, Eq.(\ref{probability}) leads to the following result
\begin{eqnarray}
&&P(\nu_{\mu}\rightarrow \nu_{\tau})\simeq \frac{1}{2}4|U_{\mu 5}|^2|U_{\tau 5}|^2\left(1-\cos \Delta m^2_{51} \frac{L}{2E}\right)+ {} \\ \nonumber 
&+&\frac{1}{2}4|U_{\mu 6}|^2|U_{\tau 6}|^2\left(1-\cos \Delta m^2_{61} \frac{L}{2E}\right)+
\frac{1}{2}4U_{\mu 5}U_{\tau 5}U_{\mu 6}U_{\tau 6}\left(1-\cos \Delta m^2_{51} \frac{L}{2E}\right)
\end{eqnarray}
Now the oscillations are also characterized by a new scale of masses, namely $\Delta m^2_{16}$ that didn't appear in the transitions $\nu_e\rightarrow\nu_{\mu},\nu_{\tau}$. 
As a rough approximation we consider just the dominant term ($\sim U_{\mu 6} U_{\tau 6}$) which implies large mixing at the scale $\Delta m^2_{16}$, that is
\begin{equation}
4|U_{\mu 6} U_{\tau 6}|^2=1, \qquad \Delta m^2_{16}\simeq \left(\frac{37}{4}\right)^2 \frac{v_R^2}{s_M}
\end{equation}
Using the estimate value for $s_M$, we found
\begin{equation}
\Delta m^2_{16} \simeq 10^{-3}\;eV^2.
\end{equation}
The recent data on atmospheric neutrino by Super-Kamiokande show that the origin of the zenith angle dependence of the neutrino flux is due to oscillations between $\nu_{\mu}$ and $\nu_{\tau}$. The data is consistent with maximal $\nu_{\mu}$ and $\nu_{\tau}$ mixing at a square mass difference scale $\Delta m^2_{atm}\simeq 10^{-3}\;eV^2$. Indeed, the preferable values of mass and mixing parameters are
\begin{equation}
\sin^22\theta_{atm}=1.0, \qquad \Delta m^2_{atm}=3.5\times 10^{-3}\;eV^2.
\end{equation}
\section{Phenomenology}
In order to analyze some phenomenological consequences of the model we'll work out the interaction Lagrangian.  We will see that the standard model results are safely recovered  at the Fermi scale and that the connection between the left and right sectors appears at the breaking scale of the new gauge group $SU(2)_R$ where non negligible effects, involving a new neutral current, are predicted.

As done elsewhere \cite{Ceron}, grouping all fermions of a given electric charge and a given helicity $(h=L,R)$ in a vector column $\psi_h=(\psi_O ,\psi_E)^t_h$ of $n$ ordinary $(O)$ and $m$ exotic $(E)$ gauge eingenstates, the interaction Lagrangian for the neutral current is simply written as
\begin{equation}\label{interaction}
\mathcal{L}^{nc}=\sum_h\overline{\psi}_h\gamma^{\mu}\left(g_LT_L^3,g_RT_R^3,g\frac{Y}{2}\right)\psi_h\left(\begin{array}{c}
W_L^3\\
W_R^3\\
B\end{array}\right),
\end{equation}
or, in terms of the physical neutral vector bosons $(Z,Z',A)$
\begin{equation}
\mathcal{L}^{nc}=\sum_h\overline{\psi}_h\gamma^{\mu}R^t\left(g_LT_L^3,g_RT_R^3,g\frac{Y}{2}\right)\psi_hR\left(\begin{array}{c}
Z\\
Z'\\
A\end{array}\right) .
\end{equation}
$R$ is a $3\times 3$ matrix representation of the orthogonal transformation which connects the weak ($W^3_{L{\mu}}$, $W^3_{R{\mu}}, B_{\mu}$) and mass eigenstates basis ($Z_{\mu},Z'_{\mu},A_{\mu}$). In its standard form, 
\begin{equation}\label{standardR}
R=\left(\begin{array}{ccc}
c_{\theta_w}c_{\alpha}&c_{\theta_w}s_{\alpha}&s_{\theta_W}\\
-s_{\alpha}c_{\beta}-c_{\alpha}s_{\theta_W}s_{\beta}&c_{\beta}c_{\alpha}-s_{\alpha}s_{\theta_W}s_{\beta}&s_{\beta}c_{\theta_W}\\
s_{\alpha}s_{\beta}-c_{\alpha}s_{\theta_W}c_{\beta}&-s_{\beta}c_{\alpha}-s_{\alpha}s_{\theta_W}c_{\beta}&c_{\beta}c_{\theta_W}
\end{array}\right)
\end{equation} 
where $\theta_{W}$, $\alpha$ and $\beta$ are the mixing angles between the $Z-A$, $Z-Z'$ and $Z'-A$ bosons.

By direct calculation from the neutral bosons mass matrix one can obtain an analytic expression for $R$ in powers of $w=v_L/v_R$
\begin{equation}\label{explicitR}
R=\left(\begin{array}{ccc}
\frac{g_L(g_R^2+g^2)^{1/2}}{\Delta^{1/2}}+O(w^4)&\frac{g_Lg^2}{(g_R^2+g^2)^{3/2}}w^2&\frac{g_Rg}{\Delta^{1/2}}\\
-\frac{g^2g_R}{\Delta^{1/2}(g_R^2+g^2)^{1/2}}-\frac{g_Rg^2\Delta^{1/2}}{(g^2+g_R^2)^{5/2}}w^2&\frac{g_R}{(g_R^2+g^2)^2}-\frac{g_Rg^4}{(g_R^2+g^2)^{5/2}}w^2&\frac{g_Lg}{\Delta^{1/2}} \\
-\frac{g^2g_R}{\Delta^{1/2}(g_R^2+g^2)^{1/2}}+\frac{g^3\Delta^{1/2}}{(g^2+g_R^2)^{5/2}}w^2& -\frac{g}{(g_R^2+g^2)^2}-\frac{g_R^2g^3}{(g_R^2+g^2)^{5/2}}w^2&\frac{g_Rg_L}{\Delta^{1/2}}\\
\end{array}\right)
\end{equation}
with $\Delta=g_L^2g_R^2+g_L^2g^2+g_R^2g^2$.

In the limit $w=0$, which corresponds to no mixing between $Z-Z'$ (or $\alpha=0$), one recovers the standard model case.

The following identities arise by comparying Eqs.(\ref{standardR}) and (\ref{explicitR}),  
\begin{equation}
\sin^2\theta_W=\frac{g^2_Rg^2}{g_R^2g_L^2+g_R^2g^2+g_L^2g^2}, \qquad \sin^2\beta=\frac{g^2}{g_R^2+g^2}.
\end{equation}

Expressed in terms of the rotation angles, the neutral currents in (\ref{interaction}) coupled to the massive vector bosons $Z$ and $Z'$ are respectively,
\begin{eqnarray}
J_{\mu}&=&\frac{g_L}{\cos\theta_W}\gamma_{\mu}\left[(1-w^2\sin^4\beta)T_{3L}-w^2\sin^2\beta T_{3R} {}\right. \\ \nonumber
&-&{}\left. Q\sin^2\theta_W(1-w^2\frac{\sin^4\beta}{\sin^2\theta_W})\right]\\
J'_{\mu}&=&g_L\tan\theta_W\tan\beta \left[\left(1+w^2\frac{\sin^2\beta \cos^2\beta}{\sin^2 \theta_W}\right)T_{3L}+\frac{T_{3R}}{\sin^2\beta} {}\right. \\ \nonumber
&-&{}\left. Q(1+w^2\sin^2\beta \cos^2\beta)\right].
\end{eqnarray}

The corrections to the standard model neutrino NC coming from the extended group symmetry are 
\begin{eqnarray}\label{correctionsw}
\mathcal{L^{\nu ,N}}&=&-J^{\nu ,N}_{\mu}Z^{\mu}-J'^{\nu ,N}_{\mu}Z'^{\mu} \\ \nonumber
&=&-\frac{g_L}{2\cos\theta_W}\left[(1-w^2\sin^{4}\beta )\overline{\nu}_L\gamma^{\mu}\nu_L- w^2 \sin^2\beta \overline{N}_R\gamma^{\mu}N_R\right]Z_{\mu}  {}\\ \nonumber
&-& \frac{1}{2}g_L\tan\theta_W\tan\beta\left[\left(1+w^2\frac{\sin^2\beta\cos^2\beta}{\sin^2\theta_W}\right)\overline{\nu}_L\gamma^{\mu}\nu_L 
+\frac{1}{\sin^2\beta}\overline{N}_R\gamma^{\mu}N_R \right]Z'_{\mu}, {}\\
\end{eqnarray}
or, in terms of the Majorana fields defined in (\ref{majoranafields}), 
\begin{eqnarray}
\mathcal{L^{\nu ,N}}&=&-\frac{g_L}{2\cos\theta_W}\left[(1-w^2\sin^{4}\beta )\sum_{i}^n \overline{\chi_{\nu i}}\gamma_{\mu}\frac{(1-\gamma_5)}{2}\chi_{\nu i} \right. + {} \\ \nonumber
&-& \left. w^2\sin^2\beta \sum_{i}^n \overline{w_{Ni}}\gamma_{\mu}\frac{(1+\gamma_5)}{2}w_{Ni} \right] Z^{\mu} {}\\ \nonumber 
&-& \frac{1}{2}g_L\tan\theta_W\tan\beta\left[(1+w^2\frac{\sin^2\beta\cos^2\beta}{\sin^2\theta_W})\sum_{i}^{n}\overline{\chi_{\nu i}}\gamma_{\mu}\frac{(1-\gamma_5)}{2}\chi_{\nu i} \right. + {}\\ \nonumber
&+& \left. \frac{1}{\sin^{2}\beta}\sum_{i}^{n} \overline{w_{Ni}}\gamma_{\mu}\frac{(1+\gamma_5)}{2}w_{Ni} \right] Z'^{\mu}
\end{eqnarray}

In order to express the neutral currents in terms of mass eigenstates one has to use the transformation relation (\ref{orthogU}) into the interaction lagrangian (\ref{correctionsw}).
This yields
\begin{eqnarray}
\mathcal{L}&=&-J^{\nu ,N}_{\mu}Z^{\mu}-J'^{\nu ,N}_{\mu}Z'^{\mu} \\ \nonumber
&=&-\frac{g_L}{2\cos\theta_W}\left[(1-w^2\sin^{4}\beta) \sum_{i=1}^3\sum_{jk}^{2n} (U_{ij}U_{ik})\overline{\eta}_ j\gamma_{\mu}\frac{(1-\gamma_5)}{2}\eta_k \right. + {} \\ \nonumber
&-& \left. w^2\sin^2\beta \sum_{i=4}^6\sum_{jk}^{2n} (U_{ij}U_{ik}) \overline{\eta}_j\gamma_{\mu}\frac{(1+\gamma_5)}{2}\eta_k \right] Z^{\mu} {}\\ \nonumber 
&-& \frac{1}{2}g_L\tan\theta_W\tan\beta\left[\left(1+w^2\frac{\sin^2\beta\cos^2\beta}{\sin^2\theta_W}\right)\sum_{i=1}^3\sum_{jk}^{2n}(U_{ij}U_{ik})\overline{\eta}_j\gamma_{\mu}\frac{(1-\gamma_5)}{2}\eta_k \right. + {}\\ \nonumber
&+& \left. \frac{1}{\sin^{2}\beta} \sum_{i=4}^6\sum_{j,k}^{2n}(U_{ij}U_{ik})\overline{\eta}_j\gamma_{\mu}\frac{(1+\gamma_5)}{2}\eta_k \right] Z'^{\mu}
\end{eqnarray}

As a consequence of the neutral gauge bosons mixing in (\ref{standardR}), mirror neutrinos couple to the $Z$ boson and may contribute to $Z$ decay  $\Gamma_Z$. Correspondence with the experimental results may be achieved by constraining the angle $\alpha$, or equivalently, the factor $w^2$, which parametrize the $Z-Z'$ mixing. This is indeed the case for $v_R> 30 v_L$ \cite{2}.
 It should be noticed that the non standard $E_L-Z$ coupling contain a term that is not suppressed by a $w^2$ factor, namely,
\begin{eqnarray}
\mathcal{L}_{Z'}^{E_L}&=& -g_L \tan \theta_W \tan \beta \overline{E_L}\gamma_{\mu}E_L.
\end{eqnarray}
However, this contribution is excluded at energies lying in the electroweak scale due to the large charged fermion masses in the heavy sector (see Sec.3). Therefore, the standard model results are recovered in the limit $w^2<<1$.

The neutral current coupled to the massive vector boson $Z'$ contains non suppressed couplings which involves either standard or exotic neutrinos and are important to test the model at the  $SU(2)_R$ breaking scale. 

These contributions are
\begin{eqnarray}
\mathcal{L_{Z'}}&=&- \frac{1}{2}g_L\tan\theta_W\tan\beta\left[\sum_{i=1}^3\sum_{jk}^{2n}(U_{ij}U_{ik})\overline{\eta}_j\gamma_{\mu}\frac{(1-\gamma_5)}{2}\eta_k \right. + {}\\ \nonumber
&+& \left. \frac{1}{\sin^{2}\beta} \sum_{i=4}^6\sum_{j,k}^{2n}(U_{ij}U_{ik})\overline{\eta}_j\gamma_{\mu}\frac{(1+\gamma_5)}{2}\eta_k \right] Z'^{\mu}
\end{eqnarray}

The new $Z'$ gauge boson can be produced at the Large Hadron Collider with masses in the 1-4 TeV region \cite{2}. The implications of a new $Z'$ to the high precision electroweak data was studied by Erler and Langacker \cite{erler}.

\section{Conclusion}
The recent experimental reports on neutrino oscillations, suggesting non zero masses for neutrinos, are certainly the strongest indication for physics beyond the standard model. 
Enlarging the fermion spectrum by introducing mirror matter is a simple way to implement non zero neutrino masses in extended theories. 
In the present paper we saw that a consistent spectrum of neutrino masses and oscillation pattern can arise in a such a scenario, which is motivated by an underlying left-right symmetric  structure in the gauge group.  
We show that the well known $SU(2)_L\times SU(2)_R \times U(1)_Y$ theory, spontaneously broken into the standard $SU(2)_L\times U(1)$ at the mass scale $v_R\simeq 10^3\; GeV$, and supplemented by two Higgs singlets with vacuum parameters at the scales $s_D\simeq 10^{10}\; GeV$ and $s_M \simeq 10^{17}\; GeV$ reproduce the observed charged and neutral fermion masses. 

The new physics predicted by the model is consistent with the theoretical arguments and experimental results available on neutrino physics. The connection between the known leptons and their mirror states can be experimentally tested by a new neutral gauge boson present at the $TeV$ mass scale.  New Majorana neutrinos, such as those considered here, may be experimentaly  tested at the large hadron collider at CERN \cite{Almeida}.

\par
\acknowledgments { This work was partially supported by the Centro Latino Americano de Física ( CLAF) and the following Brazilian agencies: CNPq, FUJB, FAPERJ and FINEP.}

\end{document}